\documentclass[pre,preprint,showpacs,showkeys]{revtex4}
\usepackage{graphicx}
\usepackage{times}

\begin{document} 

\title{Order Reductions of ``Predictive Dynamical Systems''}
\author{
J.\ M.\ Aguirregabiria, \email{juanmari.aguirregabiria@ehu.es}
} 
\affiliation{Theoretical Physics, 
The University of the Basque Country, \\
P.~O.~Box 644,
48080 Bilbao, Spain} 

\bigskip

\begin{abstract} 
It has been recently pointed  out that dynamical systems depending on future values of
the unknowns may be useful in different areas of knowledge. We explore in this context
the extension of the concept of order reduction that has been useful with
singular and delay differential equations in electrodynamics and general relativity. 
We discuss some general properties of order reductions in this new context
and explore a method of successive approximations, which among other things is used
to check and improve the ``extrapolate
prediction'' and ``fixed rate prediction'' methods.
\end{abstract} 



\maketitle


\section{Introduction}\label{sec:intro}

T.~Ohira has recently proposed and analyzed a formalism and concrete
examples of dynamical systems governed by predictions of future states,
which he calls ``predictive dynamical systems''
\cite{Ohira}. Since initial conditions
are not sufficient to solve this kind of dynamical system 
(and make sure the solution is unique),
Ohira proposes two methods of predicting the
future values of the unknowns necessary to find (numerically) the
solution to the system: ``fixed rate prediction'' and ``extrapolate
prediction.'' Since both methods can at most provide some approximate
solution of the dynamical system, it may be interesting to explore other
methods that could eventually improve the quality of the approximation.

In classical electrodynamics and general relativity one finds singular
differential equations and delay differential equations for which the
usual physical initial conditions are not enough to compute the solution.
In this context the idea of order reduction has been useful 
\cite{Kerner,Sanz,Bel2,Bel3,Bel4,Bel1,Parker,JMA,sanedrin,sanedrinluis}.

We extend the concept of order reduction to predictive dynamical systems
in section~\ref{sec:concept}, as well as a method of successive approximations to
compute the solution. Since we lack a rigorous theory of this method,
a simple but illustrative example is analyzed in section~\ref{sec:example}. 
Numerical results
are discussed in section~\ref{sec:numer} for the same dynamical systems
discussed in \cite{Ohira}.    In section~\ref{sec:higher} we briefly
explore higher order reductions and compare again our numerical results with those
of Ohira's.

\section{Order reductions of predictive dynamical systems}\label{sec:concept}

Although our definition and results can be readily extended to
continuous dynamical systems, for simplicity we will consider only 
discrete dynamical systems in which a physical quantity $x$ is defined
only for integer values $n=0,\ 1,\ 2,\ldots$ of ``time'' according
to a law in the form
\begin{equation}\label{eq:discrete}
x_{n+1}=M\left(x_n,x_{n+p}\right),
\end{equation}
with some advance $p= 1,\ 2,\ldots$
It is obvious that an initial condition $x_0$ for, say, $n=0$ is not 
enough to predict the future. Even if one could solve (\ref{eq:discrete})
for $x_{n+p}$, the resulting dynamical system
\begin{equation}\label{eq:discretesol}
x_{n+p}=N\left(x_n,x_{n+1}\right)
\end{equation}
would require specifying $p$ initial conditions: $x_0,\ x_1,\ldots,\ x_{p-1}$.
In consequence more assumptions are necessary to solve (\ref{eq:discrete}).
In Ohira's ``fixed rate prediction'' method \cite{Ohira}, one replaces
$x_{n+p}$ on the right hand side of (\ref{eq:discrete}) by 
$x_n+p\left(x_n-x_{n-1}\right)$, which still needs some additional 
initial condition for $n=-1$ or another suitable assumption.  In the 
``extrapolate prediction'' method one would substitute 
for $x_{n+p}$ the value obtained by applying $p$ times
to $x_n$ the map (\ref{eq:discrete}) with $p=0$. It is clear that, in general,
both methods will provide at most approximations to a solution of
the original dynamical system.

The idea behind order reductions is that  (\ref{eq:discrete})
is not the true evolution equation but only a necessary condition
every solution to the actual (unknown) dynamical system must satisfy.
If the true dynamical system is a deterministic one 
in the form
 \begin{equation}\label{eq:orderred}
x_{n+1} =F\left(x_n\right),
 \end{equation}
knowledge of the future is only necessary because our incomplete theory
did not led us to (\ref{eq:orderred}) but only to a less restrictive
condition in the form (\ref{eq:discrete}). Since the latter must
be satisfied by every solution to (\ref{eq:orderred}), we have the following
condition for the unknown $F$:
\begin{equation}\label{eq:condition}
F(x)=M\left(x,F^p(x)\right),\qquad  F^p\equiv\stackrel{p\ \mathrm{ times}}{\overbrace{F\circ F\circ\cdots\circ F}}.
\end{equation}

Of course, in general, one cannot solve (\ref{eq:condition}) for $F$
(this is the reason it is unknown) but one can try finding good
approximations by different methods. In some cases there is a small
parameter in the problem, so that the natural way would be to try
Taylor expansions with respect to that parameter. But we are here going to
explore a general method of successive approximations which have proved
useful with singular al delay differential equations \cite{JMA,sanedrin,sanedrinluis}.

We will construct a succession of approximations $F_0,F_1,\ldots$
defined by
\begin{equation}\label{eq:succ}
F_{m+1}(x)=M\left(x,F_m^p(x)\right)
\end{equation}
along with some suitable initial  $F_0(x)$. It is clear that if the succession
is convergent, its limit $F(x)\equiv\lim_{m\to\infty}F_m(x)$ is a solution of (\ref{eq:condition}).
One obvious choice for the initial condition is 
\begin{equation}\label{eq:xi0}
F_0(x)=M\left(x,x\right),
\end{equation}
in which case $F_1\left(x_n\right)$ is the value $x_{n+1}$ obtained by means of 
Ohira's ``extrapolate prediction;'' but, although the limit $F\left(x_n\right)$ will
be unattainable in practice, the approximation can be improved by computing successive $F_m\left(x_n\right)$ 
until  $\left|F_{m+1}\left(x_n\right)-F_m\left(x_n\right)\right|$ is below some tolerance value.
However, we will see later that different initial conditions may change dramatically the
convergence rate (moreover, the may lead to different order reductions),
so that in practice some additional criterion must be used (for instance,
in electrodynamics one can use the limit in which the charge vanishes to select the right
order reduction).

The  problem of the existence 
of the limit $F$ is here posed in too general grounds to have an answer. Instead of 
that we will consider an artificial but illustrative problem.

\section{A linear example}\label{sec:example}

Let us first consider the discrete dynamical system
\begin{equation}\label{eq:linear}
x_{n+1}=ax_n+bx_{n+1},\quad (a\ne0,\ b\ne0,1).
\end{equation}
Of course, this can be written as
\begin{equation}\label{eq:explicit}
x_{n+1}=\alpha x_n,\qquad\alpha\equiv\frac{a}{1-b};
\end{equation}
but let us pretend we do not know that and want to solve (\ref{eq:linear}) by
the method of successive approximations. It is easy to see that 
for $F_0(x)=(a+b)x$, which corresponds to (\ref{eq:xi0}), or for any
$F_0(x)=\alpha_0x$ with constant $\alpha_0$, we have
\begin{equation}\label{eq:xiam}
F_m(x)=\alpha_mx
\end{equation}
with
\begin{equation}\label{eq:recur}
\alpha_{m+1}=a+b\alpha_m.
\end{equation}
Since $\left(\alpha-\alpha_{m+1}\right)=b\left(\alpha-\alpha_{m}\right)$, whatever $\alpha_0$ is, 
the recurrence (\ref{eq:recur})
will converge (to $\alpha$) if and only if $|b|<1$. In consequence, in this example the 
method of successive approximations will converge (to the right dynamical system) when $|b|<1$ and
diverge for $|b|>1$. One cannot expect the method to be convergent always, but the example suggests that 
(as is often the case in electrodynamics \cite{JMA,sanedrin,sanedrinluis}) it may work
if some parameter in the theory is small enough. 

The following example is
\begin{equation}\label{eq:linear2}
x_{n+1}=ax_n+bx_{n+2},\quad (a\ne0,\ b\ne0,\ -1-a).
\end{equation}
Also in this case we can solve for $x_{n+2}$ to obtain a two-point recurrence which
needs two initial conditions (say $x_{-1}$ and $x_0$). Instead, we seek an order reduction (\ref{eq:orderred})
which only requires one initial condition and must satisfy
\begin{equation}
F(x)=ax+bF(F(x)).
\end{equation}
For $4ab\le1$ this functional equation has, at least, the following
two linear solutions:
\begin{equation}
F(x)=\alpha x,\qquad \alpha=\alpha_\pm\equiv\frac{1\pm\sqrt{1-4ab}}{2b}.
\end{equation}

Starting from any $F_0(x)=\alpha_0x$ with constant $\alpha_0$ we get again (\ref{eq:xiam}) with
\begin{equation}\label{eq:recur2}
\alpha_{m+1}=a+b\alpha_m^2.
\end{equation}
But this quadratic map is just the logistic map whose properties have been 
explored in depth in chaos theory \cite{Ott}. For this reason it is easy to prove that
$\alpha_m$ will converge to $\alpha_-$ for any parameter values such that $-3<4ab<1$ provided
the initial condition is choosed so that
\begin{equation}
\left|\alpha_0\right|\le\frac{1+\sqrt{1-4ab}}{2|b|}.
\end{equation}
This is the case for $\alpha_0=a+b$ ---which correspond to (\ref{eq:xi0})--- for $|a+b|$ small enough.
For other initial conditions or parameters $\alpha_m$ may go to infinity, approach
a  cycle  of any period or change chaotically. Again we see that the method could work
for small parameter values, but also that it could never converge to the right solution ($F(x)=\alpha_+x$,
for instance), in which case other methods should be tried (maybe an appropriate
series expansion, or a numerical method to solve (\ref{eq:discrete}) for $x_{n+p}$).

\section{Numerical results}\label{sec:numer}

\begin{figure}
\begin{center}
\includegraphics[width=.9\textwidth]{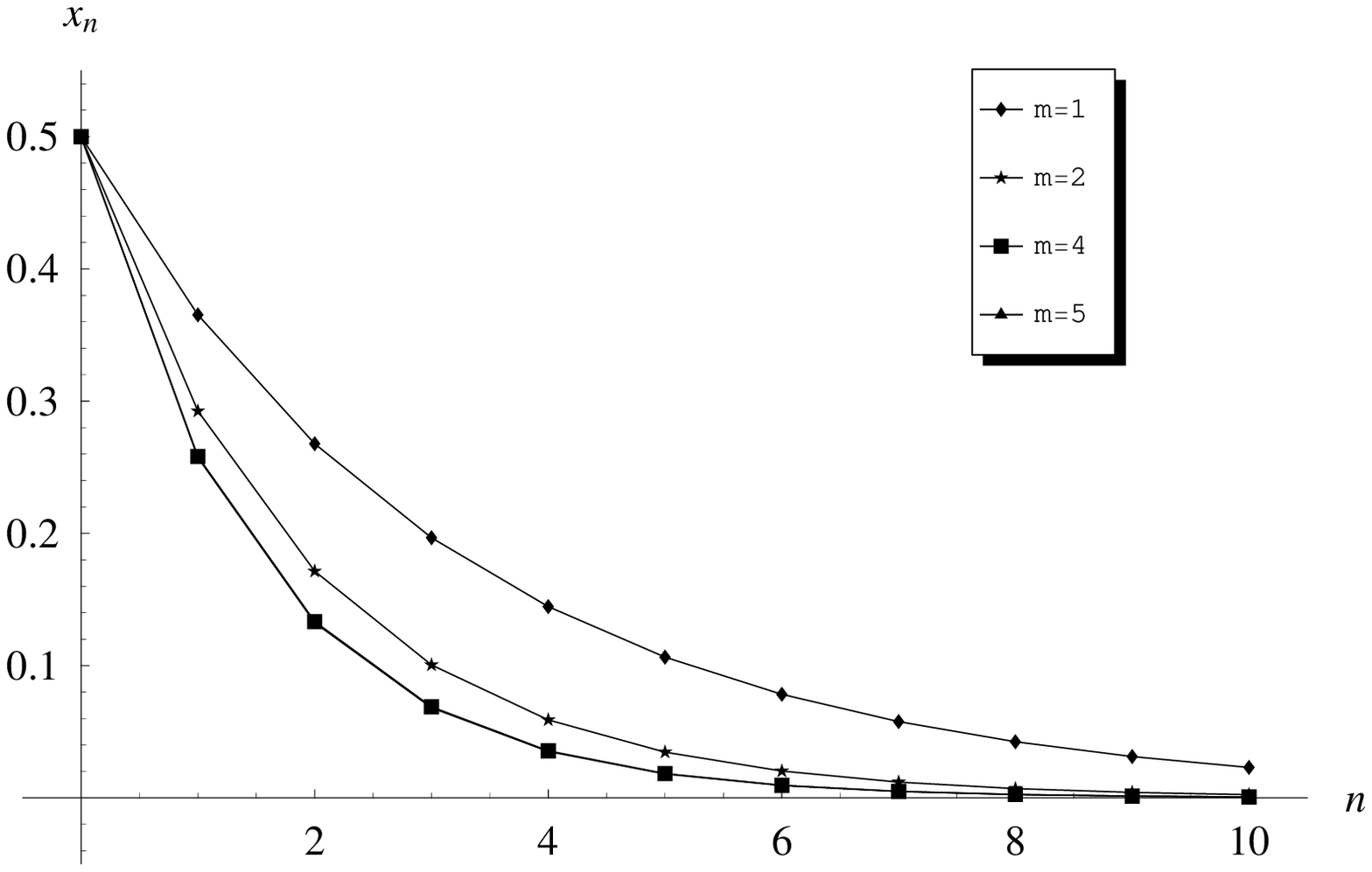}
\end{center}
\caption{$x_0,\ x_1,\ldots,x_{10}$ obtained with
$F_1$, $F_2$, $F_4$ and $F_5$, in the case
of the ``sigmoid function.''\label{fig1}} 
\end{figure}

\begin{figure}
\begin{center}
\includegraphics[width=.9\textwidth]{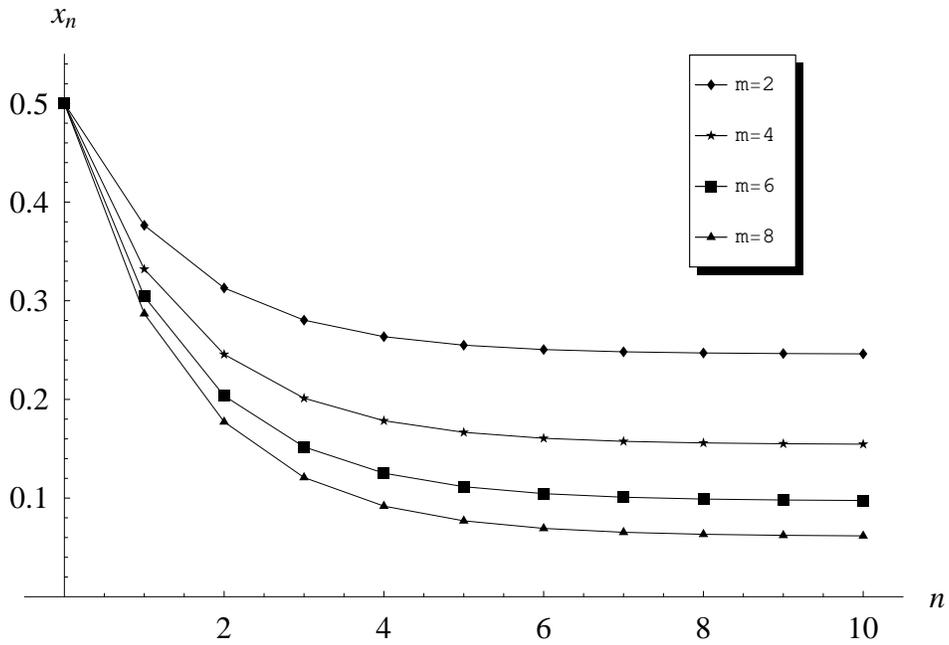}
\end{center}
\caption{Same as Fig.~\ref{fig1} but with $F_0(x)=M\left(x,x_0\right)$.\label{fig2}} 
\end{figure}

Successive approximations to the order reduction can be
numerically computed in any programming language.
For instance, the Mathematica code \cite{Math} in Table~\ref{table1}
will compute and display $x_n$ (for $n=0,1,\ldots,10$)
by using the second approximation $F_2$, in the case
of the ``sigmoid function'' discussed in reference \cite{Ohira}:
\begin{equation}
M(x,y)=(1-\mu)x+\frac2{1+e^{-\beta y}}-1.
\end{equation}
We have use that code for Figure~\ref{fig1}, where the
values $x_0,\ x_1,\ldots,x_{10}$ obtained with
$F_1$, $F_2$, $F_4$ and $F_5$ are displayed for $\mu=0.5$, $\beta=0.8$,
 $p=5$ and initial guess (\ref{eq:xi0}). The dots in the upper polygonal have been 
computed with $F_1$ and, thus,  are the same obtained by Ohira's
``extrapolate prediction.'' We can see  there is room for improvement,
for the values with $F_2$ are rather smaller, while those
obtained with $F_4$ and $F_5$ are indistinguishable in the figure,
proving they are very near those one would obtain with the limit $F$.
We can see in Figure~\ref{fig2} the importance of a good guess for $F_0$: 
selecting $F_0(x)=M\left(x,x_0\right)$
leads to a much slower convergence and even $F_8$ is not a good approximation.

\begin{figure}
\begin{center}
\includegraphics[width=.9\textwidth]{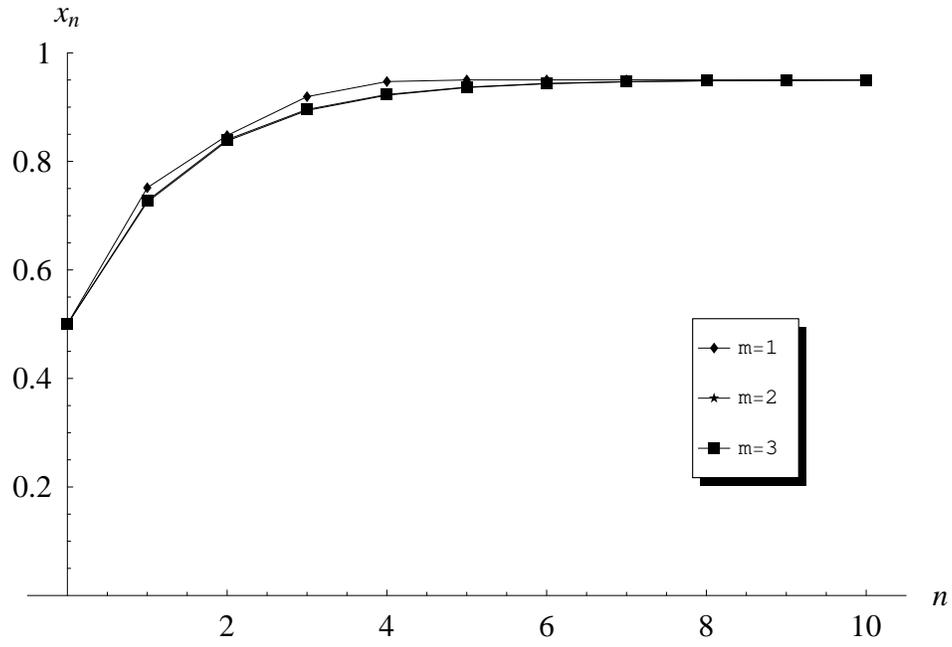}
\end{center}
\caption{$x_0,\ x_1,\ldots,x_{10}$ obtained with
$F_1$, $F_2$ and $F_3$, in the case
of the ``Mackey-Glass function.''\label{fig3}} 
\end{figure}

In Figure~\ref{fig3} one can see that convergence is faster
for the ``Mackey-Glass function'' of reference \cite{Ohira},
\begin{equation}
M(x,y)=(1-\mu)x+\frac{\beta y}{1+y^s},
\end{equation}
with $\mu=0.5$, $\beta=0.8$, $s=10$, $p=5$ and initial guess (\ref{eq:xi0}):
solutions with $F_2$ and $F_3$ are already very close.

One can also have the program compute at each step $x_n$ successive
approximations $F_m\left(x_n\right)$ until the difference between
two consecutive approximations is below some maximum
relative error, which is called \texttt{tol} in the code
in Table~\ref{table2}
for the ``Mackey-Glass function'' of reference \cite{Ohira}. For
more complex calculations this code can (must) be improved
in many ways, including a better storage management (here every
computed value is stored) and using a compiled programming language.

\section{Higher order reductions}\label{sec:higher}

\begin{figure}
\begin{center}
\includegraphics[width=.9\textwidth]{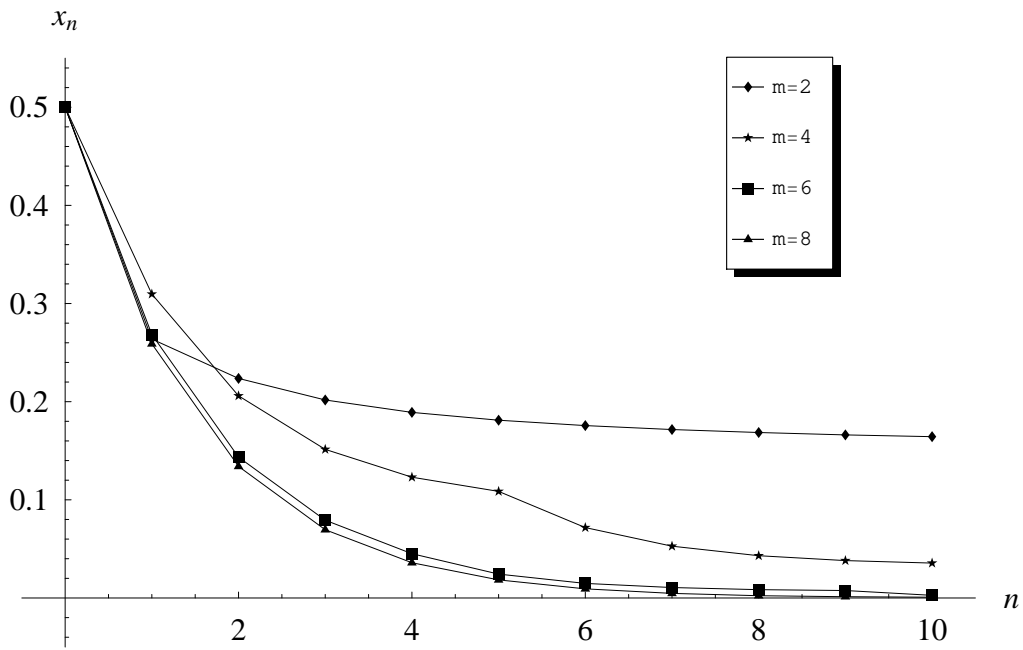}
\end{center}
\caption{$x_0,\ x_1,\ldots,x_{10}$ obtained with
$G_2$, $G_4$, $G_6$ and $G_8$, in the case
of the ``sigmoid function.''\label{fig4}} 
\end{figure}

To keep things simple we have reduced (\ref{eq:discrete}) to the first-order dynamical system
(\ref{eq:orderred}), which only needs $x_0$ to identify each solution. 
In some cases we might have theoretical reasons to think that the true dynamical
system is of second order,
 \begin{equation}\label{eq:orderred2}
x_{n+1} =G\left(x_n,x_{n-1}\right),
 \end{equation}
with
\begin{equation}\label{eq:condition2}
G(x,y)=M\left(x,G^{(p)}(x,y)\right),
\end{equation}
and
\begin{equation}
G^{(0)}(x,y)\equiv x,\quad G^{(1)}(x,y)\equiv G(x,y),\quad G^{(p+1)}(x,y)\equiv G\left(G^{(p)}(x,y),G^{(p-1)}(x,y)\right),
\end{equation}
so that $x_{-1}$ and $x_0$ must be specified. Notice that in the corresponding
scheme of successive approximations,
\begin{equation}
G_{m+1}(x,y)=M\left(x,G_m^{(p)}(x,y)\right),
\end{equation}
one could use Ohira's ``fixed rate prediction'' \cite{Ohira} to provide the following
 starting guess:
\begin{equation}\label{eq:fixed}
G_0(x,y)=M\left(x,x+p(x-y)\right).
\end{equation}
We have used the code in Table~\ref{table3} to compute the results in Figure~\ref{fig4}, 
where the results for $G_2$, $G_4$, $G_6$ and $G_8$
are displayed for $\mu=0.5$, $\beta=0.8$,
 $p=5$, $x_0=x_{-1}=0.5$ and initial guess (\ref{eq:fixed}). Wee see that the
 successive approximations converge slowly to the same solution displayed in Figure~\ref{fig1}:
 in particular this means that in this example (\ref{eq:orderred}) is also an order reduction of (\ref{eq:orderred2}).

\section{Final comments}\label{sec:concl}

We have extended the concept of order reduction to predictive dynamical systems
and discussed some examples in which it can be used to construct
good approximations to exact solutions of those systems. In particular,
we have shown that a method of successive approximations may be
used to check and improve the accuracy of Ohira's extrapolate prediction \cite{Ohira}.
We are not claiming that the method will work always,
but that, as happens with singular and delay differential equations, 
there may be interesting cases in which it can be used to
construct solutions to predictive dynamical systems. Iin other cases one must have to 
resort to other approximation scheme, such as series expansions, backward integration, shooting methods,
(or a root finding routine to solve for $x_{n+p}$ at each step), etc. 

To keep things simple we have only considered discrete dynamical systems; but the concepts
explored here can be extended in an obvious way both to reductions of higher order and to
differential-difference equations of advanced type (with the meaning defined in
reference  \cite{Bellman}).

\acknowledgments
This work was supported by The University of the Basque Country
(Research Grant~9/UPV00172.310-14456/2002).


\clearpage
%
%
%

\clearpage
\begin{table}\footnotesize
\begin{verbatim}
Clear[F]                                  (* Forget previous calculation *)
F[m_,x_] := F[m,x] =
            M[x,Nest[F[m-1,#]&,x,p]]      (* Recurrence *)
F[0,x_]  := F[0,x] = M[x,x]               (* Initial guess *)
M[x_,y_] := (1-0.5)x+2/(1+Exp[-0.8 y])-1  (* Map *)
p = 5;                                    (* Advance *)
x0 = 0.5;                                 (* Initial condition *)
m = 2;                                    (* Approximation *)
ListPlot[NestList[F[m,#]&,x0,10]];        (* Plot F_m(x_n) *)
\end{verbatim}
\caption{Mathematica program to compute and display $F_m\left(x_n\right)$.\label{table1}} 
\end{table}
%
\begin{table}\footnotesize
\begin{verbatim}
Clear[F]                                  (* Forget previous calculation *)
F::"iterations" = "Too many iterations.";
F[x_] := F[x] =
         Module[{m},                      (* Successive approximations *)
                For[m = 1,
                    Abs[(F[m,x]-F[m-1,x])/(F[m,x]+0.001)] > tol,
                    m++,
                    If [m > mmax, Message[F::"iterations"]; Break[]]
                   ];
                F[m,x]
               ]
F[m_,x_] := F[m,x] =
            M[x,Nest[F[m-1,#]&,x,p]]      (* Recurrence *)
F[0,x_]  := F[0,x] = M[x,x]               (* Initial guess *)
M[x_,y_] := (1-0.5)x+0.8y/(1+y^10)        (* Map *)
p = 8;                                    (* Advance *)
x0 = 0.5;                                 (* Initial condition *)
tol = 10^-5;                              (* Maximum relative error *)
mmax = 10;                                (* Maximum value of m *)
ListPlot[NestList[F,x0,10]];              (* Plot F(x_n) *)
\end{verbatim}
\caption{Mathematica program to compute and display an approximation to $F\left(x_n\right)$.\label{table2}} 
\end{table}
%

%
\begin{table}\footnotesize
\begin{verbatim}
Clear[G,x]                                     (* Forget previous calculation *)
G[0,m_,x_,y_] := x                             (* Recurrence *)
G[p_,m_,x_,y_] := G[p,m,x,y] = G[1,m,G[p-1,m,x,y],G[p-2,m,x,y]]
G[1,m_,x_,y_] := G[1,m,x,y] = M[x,G[p,m-1,x,y]]      
G[1,0,x_,y_] := G[1,0,x,y] = M[x,x+p(x-y)]     (* Initial guess *)
M[x_,y_] := (1-0.5)x+2/(1+Exp[-0.8 y])-1       (* Map *)
p = 5;                                         (* Advance *)
x[m_,-1] := 0.5;                               (* Initial conditions *)
x[m_,0] = 0.5;
x[m_,n_] := x[m,n] = G[1,m,x[m,n-1],x[m,n-2]]  (* Order reduction *)
m = 1;                                         (* Approximation *)
ListPlot[Table[x[m,n],{n,0,50}]];              (* Plot solution *) 
\end{verbatim}
\caption{Mathematica program to compute and display  $G_m\left(x_n,x_{n-1}\right)$.\label{table3}} 
\end{table}

\end{document}